\documentclass[aps,prl,twocolumn,superscriptaddress,groupedaddress,showpacs]{revtex4-1}
\usepackage{graphicx,amsmath,color}
\usepackage{setspace}
\def\k{\kappa}
\def\w{\omega}

\def\e{\epsilon}
\def\ve{\varepsilon}
\def\<{\langle}
\def\>{\rangle}

\begin{document}

\title{Stochastic approach to phonon-assisted optical absorption}

\author{Marios Zacharias}
\author{Christopher E. Patrick}
\altaffiliation{Current address:
Center for Atomic-Scale Materials Design (CAMD), 
Department of Physics, Technical University of Denmark.}
\author{Feliciano Giustino}
\affiliation{Department of Materials, University of Oxford, Parks Road, Oxford OX1 3PH, United Kingdom}

\date{\today}

\begin{abstract}
We develop a first-principles theory of phonon-assisted optical absorption in semiconductors
and insulators which incorporates the temperature dependence of the electronic structure.
We show that the Hall-Bardeen-Blatt theory of indirect optical absorption and the Allen-Heine theory
of temperature-dependent band structures can be derived from the present formalism by retaining
only one-phonon processes. We demonstrate this method by calculating the optical absorption coefficient
of silicon using an importance sampling Monte Carlo scheme, and we obtain temperature-dependent 
lineshapes and band gaps in good agreement with experiment. The present approach opens the 
way to predictive calculations of the optical properties of solids at finite temperature.
\end{abstract}

\maketitle

In semiconductors and insulators exhibiting indirect band gaps the optical transitions near 
the fundamental edge require the absorption or emission of phonons in order to fulfill 
the crystal momentum selection rule. This mechanism is discussed in every introduction to 
solid state physics \cite{Kittel1976,Ashcroft1976}. The theory of phonon-assisted indirect 
optical transitions was developed by Hall, Bardeen, and Blatt (HBB) \cite{hbb,Bassani_book},
and forms the basis for our current understanding of phonon-assisted optical processes.

Despite the popularity of the HBB theory, only very recently was this formalism combined 
successfully with first-principles density-functional theory calculations \cite{Noffsinger} powered 
by Wannier interpolation \cite{Giustino2007,Marzari2012}. The work of Ref.~\onlinecite{Noffsinger} 
stands as the most sophisticated calculation of indirect optical absorption available today, 
yet it is not entirely parameter-free since an empirical shift of the absorption onset at each 
temperature was needed in order achieve agreement with experiment. This correction was unavoidable 
because the HBB theory does not take into account the temperature dependence of band structures.  

A consistent theory of temperature-dependent band structures was developed by Allen and Heine 
(AH) \cite{Allen,Cardona_1}. In recent years this approach was successfully demonstrated 
and improved within the framework of first-principles density-functional theory calculations 
\cite{Marini_2008,FG_diamond, Cannuccia2011,Antonius2014}.
Given these recent advances it is natural to ask whether the HBB theory of indirect absorption 
and the AH theory of temperature-dependent band structures could be combined in a more general 
formalism, in view of fully predictive calculations of phonon-assisted optical processes at finite 
temperature.

In this manuscript we show that the quasiclassical method introduced by Williams \cite{Williams}
and Lax \cite{Lax} (WL) provides a unified framework for calculating optical absorption spectra 
of solids, including phonon-assisted absorption and electron-phonon renormalization on the same 
footing. Indeed we show that the HBB and AH theories can be derived from the WL formalism by 
neglecting electron-phonon scattering beyond one-phonon processes. In order to demonstrate the 
power of the WL approach we calculate from first principles the phonon-assisted optical absorption 
spectrum of silicon at different temperatures using a stochastic importance sampling Monte Carlo 
method \cite{Patrick2013} and no adjustable parameters. Our calculations are in very good agreement 
with experimental spectra measured at several temperatures. We also calculate temperature-dependent 
band gaps and find good agreement with experiment. 

The premise of the conventional HBB theory is that electrons in solids experience a time-dependent 
potential which arises from the oscillatory motion of the atoms around their equilibrium positions. 
Following this premise, indirect electronic transitions are obtained within time-dependent perturbation 
theory to first order in the atomic displacements \cite{Bassani_book}. This amounts to considering 
optical transitions whereby the absorption of a photon is accompanied by the emission or absorption 
of one phonon.

At variance with the HBB point of view, in the WL approach electrons and phonons are described 
on the same footing, and optical excitations correspond to transitions between Born-Oppenheimer 
product states of electrons and quantum nuclei \cite{Williams,Lax}. The quantized final vibrational 
states are then replaced by a classical continuum, leading to an expression for the optical absorption 
which only involves the nuclear wavefunction of the initial state~\cite{Lax}. This replacement can be
justified using the adiabatic approximation. The temperature dependence is then obtained as a canonical 
average over the initial states of the system. This theory was successfully employed to explain 
the optical properties of cold lithium clusters \cite{DellaSala} and diamondoids \cite{Patrick2013}.  

  \begin{figure*}[t]
  \includegraphics[width=0.85\textwidth]{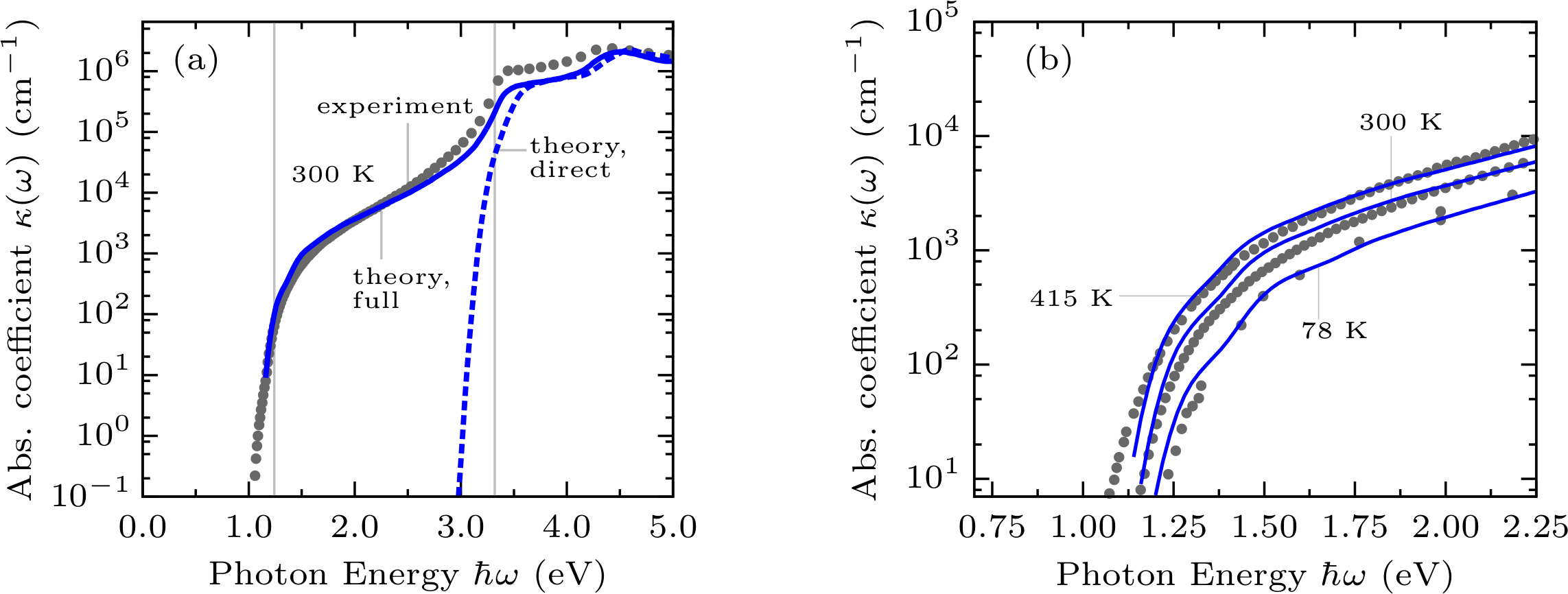}
  \caption{\label{fig1} 
  (a) The absorption coefficient of bulk silicon at 300~K: calculation with the atoms clamped at 
  their equilibrium positions (blue dashed line), calculation using the WL method [Eq.~(\ref{eq.2}), 
  blue solid line], and experimental data from Ref.~\onlinecite{M_Green3} (grey filled discs).  
  The thin vertical lines indicate the direct and indirect band gaps with nuclei in their equilibrium
  positions. (b)~Temperature dependence of the absorption coefficient of silicon: WL theory (solid lines) 
  and experimental data for 78~K \cite{Braunstein}, 300~K \cite{M_Green3}, and 415~K \cite{Weakliem} 
  (grey discs). The calculated spectra were broadened using Gaussians of width 30~meV, and truncated 
  at the smallest excitation energy in order to avoid artifacts. 
  }
  \end{figure*}

The imaginary part of the WL temperature-dependent dielectric function is given by:
  \begin{equation}\label{eq.1}
  \e_2(\w;T)  = Z^{-1} {\sum}_n \exp(-E_n /k_{\rm B}T) \< \e_2(\w;x) \>_n,
  \end{equation}
where $\w$ is the photon frequency, $k_{\rm B}$ the Boltzmann constant, and $T$ the temperature. 
$Z$ is the canonical partition function among the quantum nuclear states with energies $E_n$, and 
$\<\,\,\>_n$ stands for the expectation value taken over the $n$-th many-body nuclear wavefunction.
$\e_2(\w;x)$ denotes the imaginary part of the dielectric function evaluated with the nuclei 
clamped at the positions specified by the set of normal coordinates $\{x_\nu\}$, which 
we indicate collectively as $x$. In order to keep the notation light we label the normal modes 
of vibration and the electronic states by integer indices; accordingly the following equations will
refer to a Born-von K\'arm\'an (BvK) supercell of the crystal unit cell. An intuitive interpretation 
of Eq.~(\ref{eq.1}) is that in the adiabatic approximation the electronic and nuclear timescales 
are decoupled, and the measured absorption spectrum is described as an ensemble average over 
instantaneous absorption spectra at fixed nuclear coordinates. In the harmonic approximation 
Eq.~(\ref{eq.1}) simplifies via Mehler's formula~\cite{Watson1933}:
  \begin{equation}\label{eq.2}
  \e_2(\w;T)  = \int\! \Pi_\nu \, dx_\nu \,G\!\left[x_\nu;\<x_\nu^2\>_{_T}\right] \e_2(\w;x),
  \end{equation}
where $G[u;\sigma^2]$ is a normalized Gaussian of width $\sigma$ in the variable $u$. $\<x_\nu^2\>_{_T} 
= (2n_\nu+1) \,l_\nu^2$ represents the mean square nuclear displacement at the temperature $T$, with 
$n_\nu$ the Bose-Einstein occupation factor of the mode with energy $\hbar\w_\nu$ at the temperature 
$T$, and $l_\nu$ the corresponding zero-point amplitude \cite{CEP_FG}.

For simplicity we calculate the dielectric function within the independent-particle approximation, 
although the present formalism is general and can be used with any description of optical 
transitions at fixed nuclei. In the electric dipole approximation we have~\cite{Cardona_Book}:
  \begin{equation}\label{eq.3}
  \epsilon_2(\w;x) = \frac{2 \pi }{ m N } \frac{\w_{\rm p}^2}{\,\w^2}
      \sum_{cv} | p_{cv}^x|^2
   \delta(\ve_c^x-\ve_v^x-\hbar\w),
  \end{equation}
where $m$ is the electron mass, $\w_{\rm p}$ the plasma frequency, $N$ the number of electrons
in the unit cell, and the factor of two is for the spin degeneracy. $p_{cv}^x$ is the matrix element 
of the momentum operator along the polarization direction of the photon, taken between the valence 
and conduction Kohn-Sham states $|v^x\>$ and $|c^x\>$ with energies $\ve_v^x$ and $\ve_c^x$, 
respectively. The superscripts indicate that these states are calculated with the nuclei fixed 
in the configuration specified by the normal coordinates $x$; the same quantities evaluated 
at the equilibrium atomic positions will be denoted without superscripts. 
Equation~(\ref{eq.2}) was evaluated within density functional theory using importance sampling 
Monte Carlo integration in a BvK supercell, as described below in the Methods.

In Fig.~\ref{fig1}(a) we compare the optical absorption coefficient of silicon calculated from 
first principles using Eqs.~(\ref{eq.2}) and (\ref{eq.3}) with the experimental spectrum, both at 
300~K. The absorption coefficient was obtained as $\k(\w;T) = \w\, \e_2(\w;T)/c\, n(\w)$ where $c$ 
is the speed of light and $n(\w)$ the refractive index. The spectrum calculated with the nuclei clamped 
in their equilibrium positions [dashed blue line in Fig.~\ref{fig1}(a)] exhibits an onset around 3.3~eV, 
corresponding to the direct $\Gamma'_{25v}\rightarrow \Gamma_{15c}$ transition in silicon. The sub-gap 
absorption between 1.1--3.3~eV observed in experiments \cite{M_Green3} is completely missing in this 
calculation. At variance with this result, our WL spectrum correctly captures indirect absorption 
[solid blue line in Fig.~\ref{fig1}(a)], and exhibits very good agreement with experiment without 
any adjustable parameters. Since we are not including excitonic effects, the strength of the $E_1$ 
transition is underestimated in our calculations, as can be seen at energies around 3.3~eV in 
Fig.~\ref{fig1}(a) \cite{Benedict}. The agreement between theory and experiment in Fig.~\ref{fig1} 
remarkably extends over five~orders of magnitude.

In order to shed light on the ability of the WL theory to capture indirect optical absorption we 
express the dependence of the optical matrix elements on the atomic positions using time-independent 
perturbation theory. To first order in the atomic displacements we have:
  \begin{equation}
  p_{cv}^x  =p_{cv} + {\sum}_{n\nu}^\prime \left[ 
  \frac{p_{cn}\, g_{nv\nu}}{\ve_v-\ve_n} +\frac{g_{cn\nu}\, p_{nv}}{\ve_c-\ve_n}
  \right] \!\frac{x_\nu}{l_\nu}, \label{eq.p} 
  \end{equation}
where $g_{mn\nu} = \< m | \partial V/\partial x_\nu | n\> l_\nu$ is the electron-phonon matrix element 
associated with the Kohn-Sham potential $V$, and in the primed summation the terms $n=c$, $v$ are skipped.
The spectral range of indirect absorption corresponds to photon energies $\hbar\w < E_{\rm g}^{\rm d}$,
with $E_{\rm g}^{\rm d} =$~3.3~eV being the direct band gap of silicon. In this range direct optical 
transitions are forbidden, therefore from Eq.~(\ref{eq.p}) we have $p_{cv} = 0$. If we retain only 
one-phonon processes and neglect the dependence of the electron energies on the nuclear coordinates,
Eqs.~(\ref{eq.2})-(\ref{eq.p}) yield:
  \begin{eqnarray}\label{eq.hbb-1}
  \epsilon_2(\w;T) &= & \frac{2 \pi }{ m N } \frac{\w_{\rm p}^2}{\,\w^2} \sum_{cv\nu}
  \left|{\sum}_n^\prime \frac{p_{cn}\, g_{nv\nu}}{\ve_n-\ve_v} +\frac{g_{cn\nu}\, 
  p_{nv}}{\ve_n-\ve_v-\hbar\w} \right|^2  \nonumber \\ & \times &
  \delta\left(\ve_c-\ve_v -\hbar\w \right) (2n_\nu+1).
  \end{eqnarray}
This expression is essentially the same as given by the conventional HBB theory of indirect 
optical absorption~\cite{Bassani_book}, and employed in the first-principles calculations of 
Ref.~\onlinecite{Noffsinger}. The only difference is that the HBB theory contains phonon energies 
$\pm \hbar\w_\nu$ in the denominators and the Dirac delta functions, corresponding to phonon emission 
and absorption processes, respectively. 
In the WL approach these terms are neglected since in the adiabatic approximation $\hbar\w_\nu \ll 
\ve_c-\ve_v$. In Figure~S1~\cite{suppl} we show that the present result agrees well 
with the indirect absorption spectrum of silicon calculated using the conventional HBB theory in 
Ref.~\onlinecite{Noffsinger}.

In Fig.~\ref{fig1}(b) we compare our calculated temperature dependence of the indirect optical 
absorption lineshape of silicon with experiment. We focus on the energy range 1.1--2.3~eV where 
the effect of excitonic spectral weight transfer on the dielectric function is negligible. Our 
calculations are in good agreement with experiment. In particular the theoretical spectra capture 
both the smooth increase of the absorption coefficient with temperature, and the concurrent redshift 
of the absorption onset. We stress that the observed redshift arises naturally in our calculations, 
in contrast with the HBB theory where this effect needs to be included empirically \cite{Noffsinger}.
The slight loss of intensity near the indirect edge at the highest temperature [spectrum at 415~K 
in Fig.~\ref{fig1}(b)] results from the incomplete sampling of multi-phonon processes in our 
stochastic approach.

  \begin{figure*}[t]
  \includegraphics[width=0.85\textwidth]{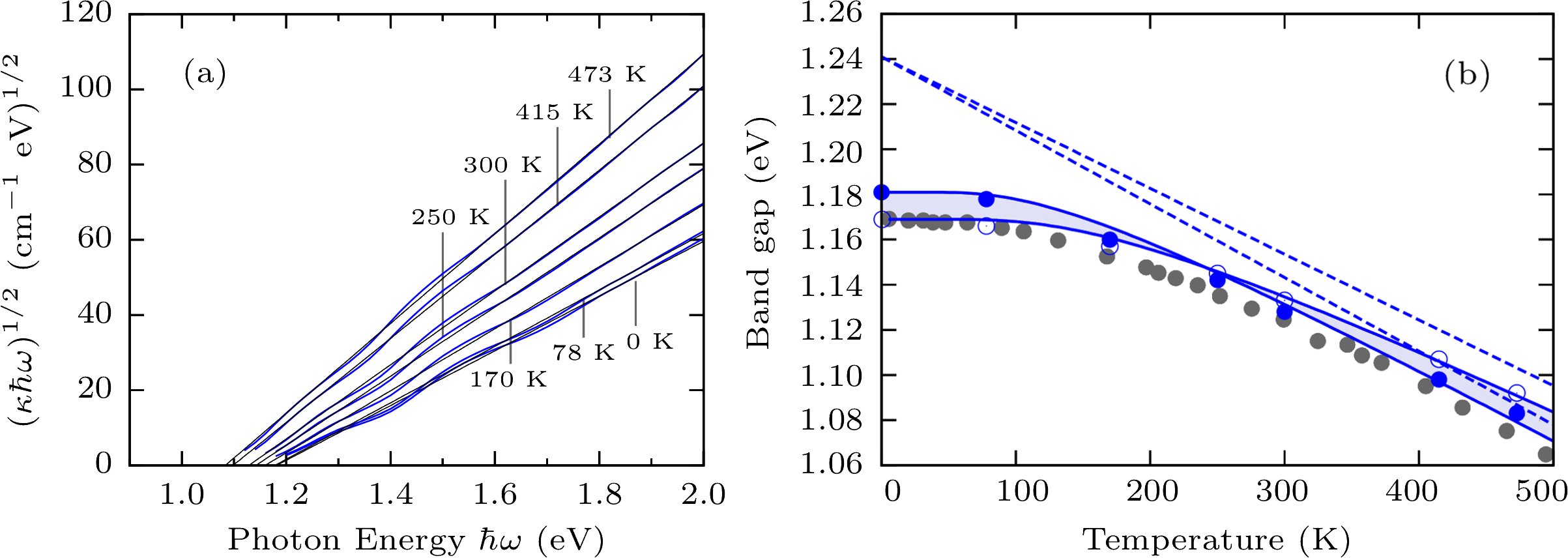}
  \caption{ \label{fig2}
  (a) Extraction of the temperature-dependent indirect band gap of silicon using lineshape analysis.
  The calculated $[\w\k(\w)]^{1/2}$ at each temperature are shown as blue lines, and the corresponding 
  linear fits as thin black lines. The intercepts of the straight lines with the horizontal axis give 
  the band gaps. The linear fits were determined in the energy range 0--2~eV. (b)~Temperature-dependent 
  indirect band gap of silicon: the band gaps extracted from the lineshape analysis in (a) using linear 
  fits in the ranges 0--2~eV and 0--1.5~eV are shown as blue filled discs and open circles, respectively. 
  Grey filled discs are experimental data from Ref.~\onlinecite{Alex}. The solid lines are single-oscillator 
  fits to the calculated data, and the dashed lines are the corresponding high-temperature asymptotes.
  Figure~S2~\cite{suppl} shows the sensitivity of the band gaps to the fitting range. The shading
  is a guide to the eye and can be taken as the uncertainty of the theoretical band gaps.
  }
  \end{figure*}

In order to understand the effect of temperature in the WL approach we note that temperature enters 
the formalism in two ways: firstly in the Bose-Einstein factors $(2n_\nu+1)$ in Eq.~(\ref{eq.hbb-1}), 
as in the conventional HBB theory. This term mainly modifies the absorption intensity. Secondly 
temperature enters in the electron-phonon renormalization of the electronic band structure, leading 
to a temperature-dependent shift of the absorption onset. The latter contribution can be analyzed 
by rewriting the energies inside the Dirac delta functions in Eq.~(\ref{eq.3}) using time-independent 
perturbation theory. The result accurate to second order in the atomic displacements reads:
  \begin{equation}
  \ve_c^x = \ve_c + \!\sum_\nu g_{cc\nu} \frac{x_\nu}{l_\nu} +
  \!\sum_{\mu\nu} \left[ {\sum}_n^\prime \frac{ g_{cn\mu} g_{nc\nu} }{\ve_c -\ve_n} + h_{c\mu\nu}
  \right] \!\!\frac{x_\mu x_\nu}{l_\mu l_\nu}, 
  \label{eq.e}
  \end{equation}
where $h_{n\mu\nu} = \< n | \partial^2 V/\partial x_\mu\partial x_\nu | n\>l_\mu l_\nu/2$ is the 
Debye-Waller electron-phonon matrix element \cite{Allen,Cardona_1}. If we evaluate the average of 
$\ve_c^x$ in Eq.~(\ref{eq.e}) following the same prescription as for the dielectric function in 
Eq.~(\ref{eq.2}) we obtain (up to third order in the displacements):
  \begin{equation}\label{eq.ah}
  \ve_c(T) = \ve_c + {\sum}_\nu \left[ {\sum}_n^\prime  
  \frac{ |g_{cn\nu}|^2  }{\ve_c -\ve_n} + h_{c\nu\nu} \right] (2n_\nu+1).
  \end{equation}
In the first term inside the square brackets we recognize the Fan (or self-energy) electron-phonon 
renormalization; the second term is the Debye-Waller renormalization \cite{Allen,Cardona_1,Giustino2007,
Marini_2008,Cannuccia2011,Antonius2014}. Both terms can be derived from a diagrammatic analysis 
by considering only one-phonon processes \cite{Marini2015}. Equation~(\ref{eq.ah}) represents precisely 
the AH theory of temperature-dependent band structures, and explains the temperature shift of the 
indirect absorption lineshapes in Fig.~\ref{fig1}(b).

From the calculated optical absorption spectra we can extract the temperature dependence of the 
indirect and direct band gaps of silicon, following the standard {\it experimental} procedure. In fact
within the HBB theory the absorption coefficient near the indirect edge goes like $\w^{-1}(\hbar\w 
-E_{\rm g}\pm\hbar\w_\nu)^2$ \cite{Cardona_Book,Bassani_book}, therefore the indirect gap $E_{\rm g}$ 
is straightforwardly extracted from a linear fit to $\w^{1/2} \k(\w)^{1/2}$. As expected 
Fig.~\ref{fig2}(a) shows that our calculated spectra follow a straight line when plotted as 
$\w^{1/2} \k(\w)^{1/2}$. The intercept of this line with the horizontal axis yields the indirect 
band gaps for each temperature, and the results are shown in Fig.~\ref{fig2}(b) for
two fitting ranges, 0--1.5~eV and 0--2~eV. Single-oscillator fits to our data using $E_{\rm g}(T)
=E_{\rm g}(0) -a_{\rm B}\{1 + 2/[ \exp(\Theta / T) -1]\}$ following Ref.~\onlinecite{Cardona_3} 
gave a zero-point renormalization of $a_{\rm B}=60$--72~meV and an effective temperature 
$\Theta=368$--494~K for the two ranges considered. These values are in good agreement with 
the experimental data 62~meV and 395~K, respectively \cite{Cardona20053}.

In Figure~S3~\cite{suppl} we show that the WL spectrum can also be used to extract the temperature
dependence of the {\it direct} band gap of silicon using standard lineshape analysis of second-derivative 
spectra. Also in this case we obtain good agreement with experiment. Overall the agreement 
between theory and experiment in Fig.~\ref{fig1}, Fig.~\ref{fig2}, and Figure~S3 provides 
strong support to the validity of the WL theory for first-principles calculations of phonon-assisted 
optical absorption spectra.

In future work it will be important to test the role of additional correction terms, such as nonadiabaticity 
\cite{Cannuccia2011}, quasiparticle corrections \cite{Antonius2014}, and anharmonicity \cite{Errea2014}. 
While these further refinements will modify the precise values of the zero-point renormalization of the 
band gap, it is expected that they will not change any of the features of the lineshapes in Fig.~\ref{fig1}.

The stochastic approach employed here is remarkably efficient in sampling the vibrational phase
space, to the point that the optical spectrum can be calculated using a {\it single configuration} 
of the nuclei (Fig.~S4 \cite{suppl}).  This is an unexpected finding and warrants separate 
investigation.  While the present method lacks the elegance of standard density-functional perturbation 
theory approaches \cite{Baroni2001}, it comes with distinctive advantages: (i) the electron-phonon 
coupling is included to all orders, (ii) the method can be used in conjunction with higher-level 
theories, such as hybrid functionals \cite{PBE0,HSE2003} and the GW/Bethe-Salpeter method \cite{Onida2002},
and (iii) the anharmonicity of the potential energy surface can be incorporated
by using the appropriate nuclear wavefunctions~\cite{Monserrat2013}.

In conclusion we have demonstrated a new theory of phonon-assisted optical absorption in solids,
based on the Williams-Lax quasiclassical approximation. This theory incorporates for the first time 
the temperature-dependent electron-phonon renormalization of the electronic structure, and enables 
calculations of optical spectra at finite temperature over a wide spectral range.
Our stochastic approach is efficient and easy to 
implement on top of any electronic structure package. The present work opens the way to systematic calculations 
of optical spectra of semiconductors and insulators at finite temperature.

{\it Methods} The calculations were performed within density functional 
theory in the local density approximation \cite{CA1980,PZ1981}, using planewave basis sets and norm-conserving 
pseudopotentials~\cite{Fuchs199967} as implemented in the {\tt Quantum ESPRESSO} suite~\cite{QE}. 
We obtained vibrational frequencies and eigenmodes via the frozen-phonon method \cite{Kunc_Martin,
Ackland}. The optical matrix elements including the non-local components of the pseudopotential
\cite{Starace_1971} were evaluated using {\tt Yambo} \cite{Marini20091392}. 
Calculations with/without the nonlocal components of the pseudopotential are compared 
in Figure~S1~\cite{suppl}. In order to address the band gap problem we used a scissor correction 
$\Delta = 0.75$~eV in all calculations, close to the GW value of Ref.~\onlinecite{Henry_FG}. 
The non-locality of the scissor operator was taken into account 
in the oscillator strengths \cite{Starace_1971} via the renormalization factors $(\ve_c-\ve_v)/(\ve_c-\ve_v
+\Delta)$, thereby ensuring that the $f$-sum rule be correctly fulfilled. 
A comparison between the absorption spectra calculated with or without the scissor correction 
is shown in Figure~S5~\cite{suppl}. We averaged over the atomic 
configurations using Importance Sampling Monte Carlo integration \cite{Patrick2013}. The estimator 
\cite{Caflisch} of $\ve_2(\w;T)$ in Eq.~(\ref{eq.2}) was obtained using configurations 
generated from a random set of normal coordinates $\{x_\nu\}$, as determined from the quantile 
function of the Gaussian distribution, $x_\nu = (2 \< x_\nu^2\>_T)^{1/2}  
\mbox{erf}^{-1}(2t-1)$ \cite{Brown}. The $t$ values (one for each normal coordinate, $0<t<1$) 
were generated via Sobol low-discrepancy sequences \cite{sobol} by skipping the first 100~steps. 
We found that 5--8 atomic configurations are enough to converge the spectra at high/low temperature, 
respectively. Figure~S4~\cite{suppl} shows that even using a single configuration the spectrum 
is already converged. The results presented in Figs.~\ref{fig1} and \ref{fig2} were 
obtained using a 8$\times$8$\times$8 BvK supercell of the silicon unit cell. We sampled the Brillouin 
zone of the supercell using 30~random points with weights determined by a Voronoi analysis \cite{Voronoi}. 
Convergence tests with respect to the supercell size and Brillouin zone sampling are shown in 
Figure~S6~\cite{suppl}. In Fig.~\ref{fig2}(a) the spectrum was calculated using a 
Gaussian broadening of 30~meV; in Figure~S4(b)~\cite{suppl} we show that even when using a broadening 
of only 1~meV the spectra remain essentially unaltered. Figure~S7~\cite{suppl} shows that
the variation of the band gap due to the thermal expansion of the lattice~\cite{Okada1984} is smaller than 
5~meV up to 500~K, and can be neglected.

The authors wish to thank E. Kioupakis, S. Ponc\'e, A. Marini, M. C\^ot\'e, and M. Ceriotti for 
many fruitful discussions. This work was supported by the Leverhulme Trust (Grant RL-2012-001),
the European Research Council (EU FP7 / ERC grant no. 239578), and the UK Engineering and Physical 
Sciences Research Council (Grant No. EP/J009857/1 and DTA support). This work used the ARCHER UK 
National Supercomputing Service via the AMSEC project, and the Advanced Research Computing facility 
of the University of Oxford.

\bibliography{references}{} 

\newpage

\onecolumngrid
\begin{center}

{\bf \Large{ Supplemental Material for \\ ``Stochastic approach to phonon-assisted optical absorption''}}

\end{center}
\newpage
\footnotesize
\setstretch{1.0}

  \begin{figure*}
  \includegraphics[width=\textwidth]{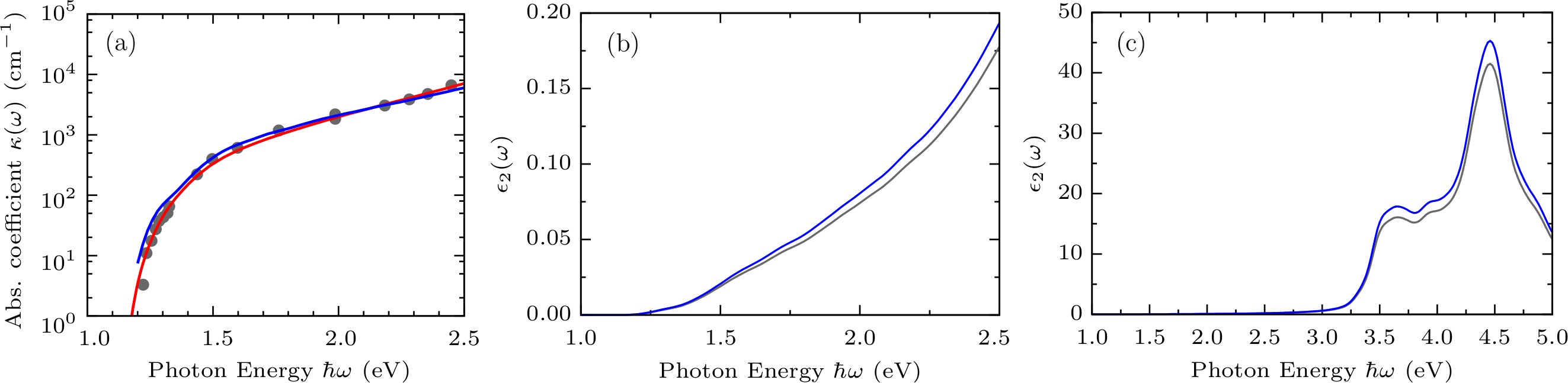}
  \end{figure*}
  \noindent FIG. S1:
  (a) Comparison between the optical absorption spectrum of silicon calculated in the present work (blue solid
  line), the spectrum calculated in Ref.~[5] using the HBB theory (red solid line),
  and the experimental data from Ref.~[19] (gray filled discs). All spectra are for
  $T=78$~K. The contributions to the optical matrix elements arising from the nonlocal components
  of the pseudopotential were not included in Ref.~[5]. For consistency in this
  figure we calculated spectra {\it without} such contributions.
  (b) and (c) Imaginary part of the dielectric function of silicon calculated in the present work at 78~K, with (blue)
  or without (gray) including the contributions to the optical matrix elements arising from the nonlocal components
  of the pseudopotential. We note that the scale in (a) is logarithmic, while in (b) and (c) we use a linear scale for clarity.
  The inclusion of the nonlocal components of the pseudopotential leads to a
  modification of the oscillator strength in the order of 10--15\%.
  We note that in Figs.~1 and 2 of the main text the non-local contributions are correctly included.
  The pseudopotential used for these calculations has the local component in the $d$ channel.

\newpage

  \begin{figure*}
  \includegraphics[width=0.6\textwidth]{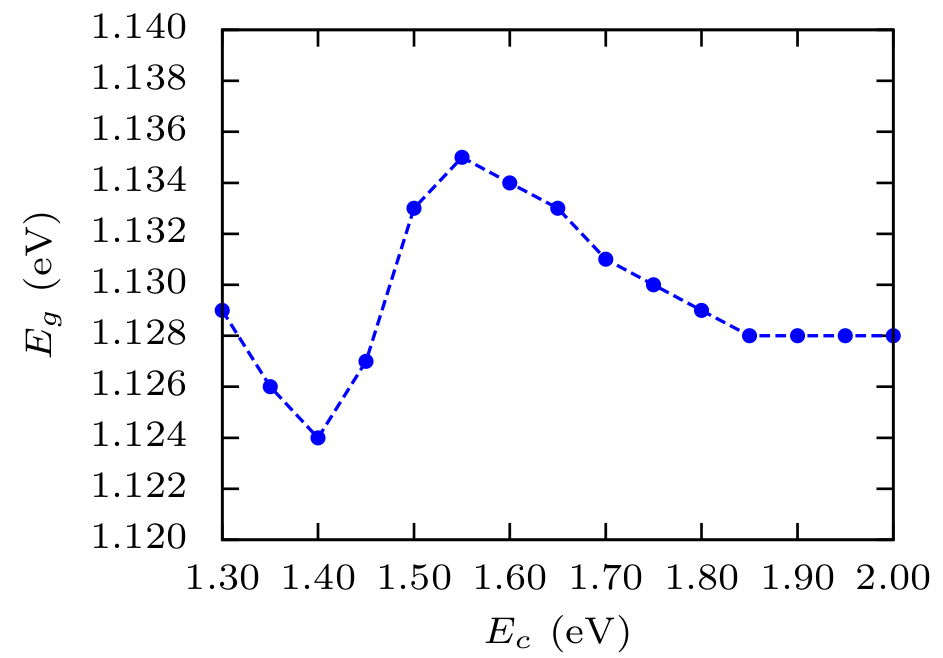}
  \end{figure*}
  \noindent FIG. S2:
  Dependence of the calculated indirect band gap of silicon on the energy cutoff $E_{\rm c}$ used
  in the linear fits of Fig.~2. The fits were performed in the energy range $[0,E_{\rm c}]$, and the
  temperature considered in this example is 300~K.
\newpage

  \begin{figure*}[t]
  \includegraphics[width=\textwidth]{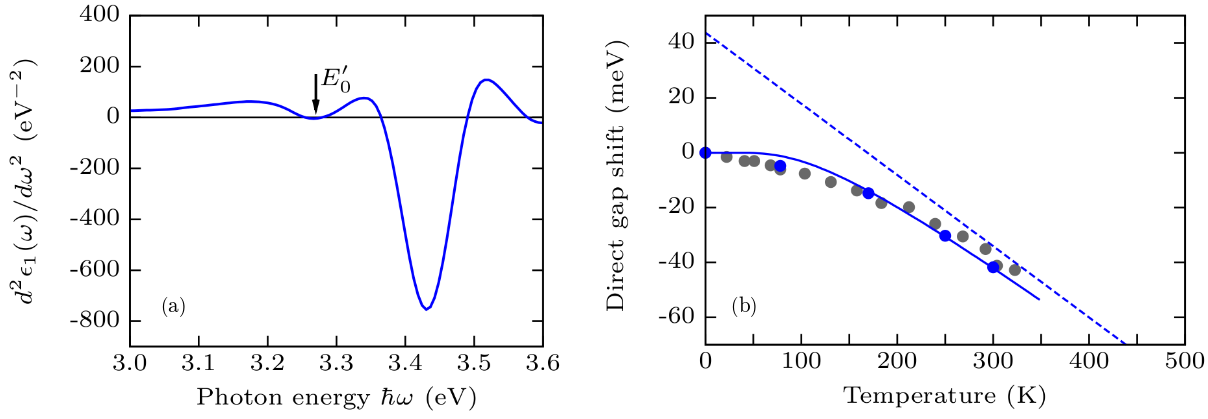}
  \end{figure*}
  \noindent FIG. S3:
  (a) Calculated second-derivative spectra of the real part of the dielectric function of silicon 
  at $T=0$~K within the WL formalism. In experiment the direct band gap of silicon is determined by 
  identifying the $E_0'$ transition with the first dip in this spectrum~[28].
  (b)~Calculated temperature dependence of the direct gap of silicon within the WL theory
  (blue filled discs), using the second-derivative method illustrated in panel (a). For comparison 
  the experimental data from Ref.~[28] are shown as gray filled discs. A single-oscillator 
  fit (see main text) describes adequately the calculated temperature dependence (thin blue line), and 
  we obtain $a_{\rm B} = 44$~meV and $\Theta = 337.5$~K, which are compatible with the experimental 
  ranges $25\pm 17$~meV and $267\pm 123$~K~[28], respectively. The high-temperature 
  asymptote is shown as a straight blue dashed line. Above 350~K the $E_0'$ and $E_1$ transitions 
  merge (both in experiments and in our calculated spectra) and the direct band gap cannot be determined 
  using this procedure.
\newpage

\begin{figure*}
  \includegraphics[width=\textwidth]{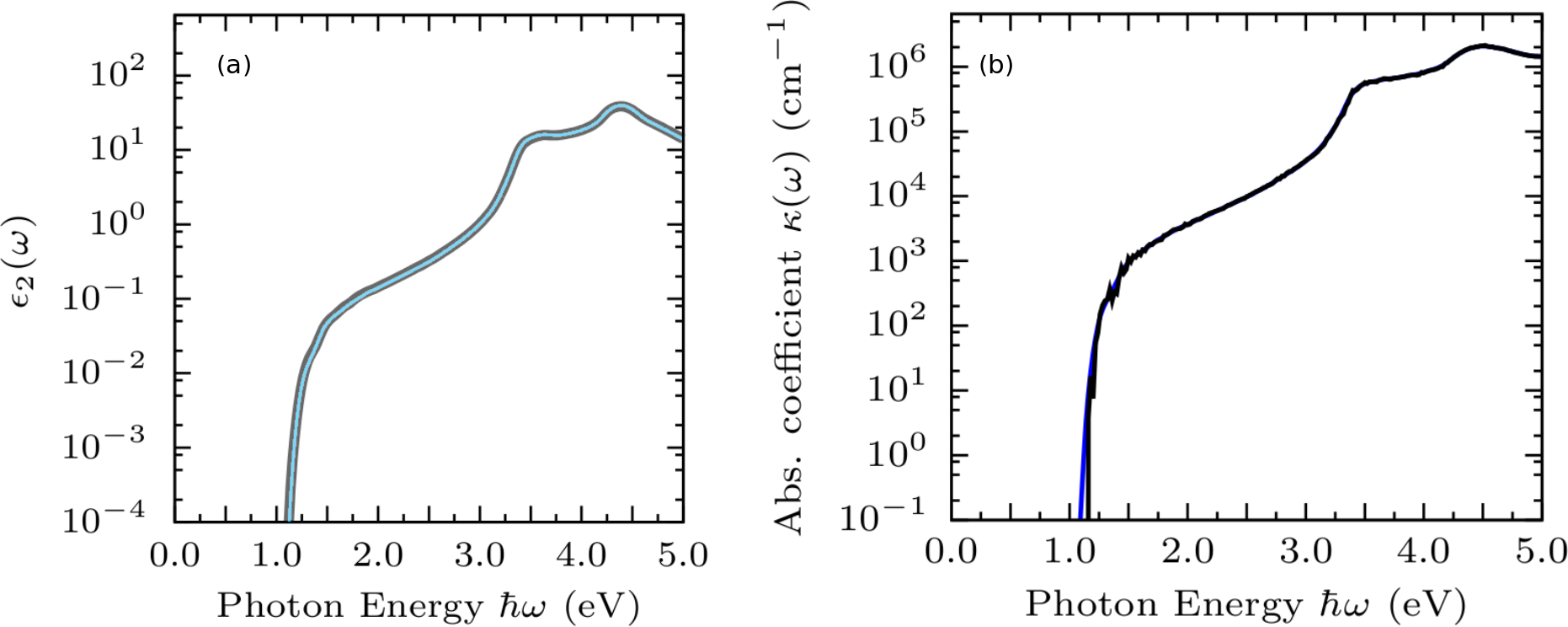}
  \end{figure*}
  \noindent FIG. S4:
  (a) Sensitivity of the calculated imaginary part of the dielectric function of silicon to the number 
  of nuclear configurations used for evaluating the Monte Carlo estimator. The thin blue line and the 
  thick gray line correspond to averages over 6 and 1 random configurations, respectively. (b)~Absorption 
  coefficient of silicon calculated using two different Gaussian broadening parameters, 30~meV (blue line) 
  and 1~meV (black line). These calculations are for a 8$\times$8$\times$8 BVK supercell and 300~K.
\newpage	  

  \begin{figure*}
  \includegraphics[width=0.55\textwidth]{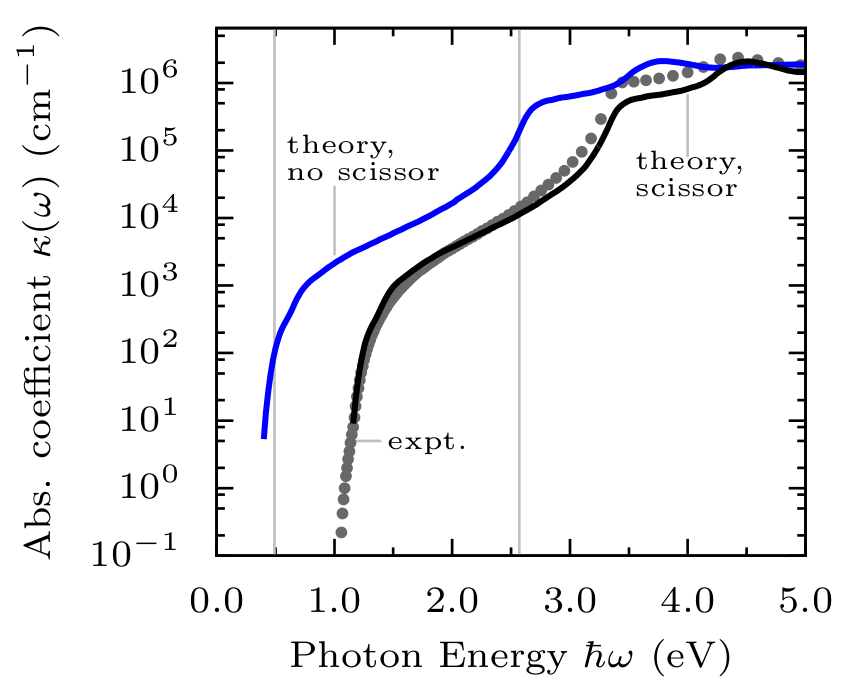}
  \end{figure*}
  \noindent FIG. S5:
Calculated absorption coefficient of bulk silicon at 300~K with (black sold line)
and without (blue solid line) the scissor operator. 
The experimental data are from Ref.~[18] (gray filled discs).
The thin vertical lines indicate the indirect and direct band gaps (without scissor),
with nuclei in their equilibrium positions.
It is seen that the scissor yields {\it essentially} a rigid blue-shift of the
entire spectrum, and does not introduce any artifacts.
We can rationalize this observation as follows.
The scissor correction modifies the imaginary part of the dielectric function in two ways:
firstly by rigidly shifting the transition energies, and secondly by renormalizing the
oscillator strengths. This latter aspect is important since the scissor operator is
a projection over the unoccupied manifold, and its non-locality needs to be taken into
account in order to obtain the correct dipole matrix element, see Ref.~[42].
Let us call the dielectric function {\it without} scissor $\e(\w)$, and that {\it with}
scissor as $\e^{\rm s}(\w)$. Based on the previous considerations, in the independent-particle
approximation the imaginary parts are related as follows:
  $\e_2^{\rm s}(\w) = (1-\Delta/\hbar\w) \e_2(\w-\Delta/\hbar)$.
From this expression it is immediate to verify that the dielectric function after scissor correction correctly
fulfills the same $f$-sum rule as the original function.
Using the above relation we find that the absorption coefficients with/without scissor
are related as follows:
  $ \k^{\rm s}(\w) = [n(\w-\Delta/\hbar)/n^{\rm s}(\w)] \k(\w-\Delta/\hbar)$.
In the present case the ratio of the refractive indices with/without scissor is found to be
approximately equal to unity in the photon energy range of interest
(0.91--1.01 for energies up to 4.5~eV). This reflects the fact
that below the direct gap the refractive index is dominated by the real part of the dielectric
function, and that there is no sharp structure in $\e_2(\w)$ below the direct gap.
As a result the absorption spectrum in the presence of scissor correction appears simply
as a blue-shifted version of the un-modified spectrum, as shown in the figure.

\newpage

  \begin{figure*}
  \includegraphics[width=\textwidth]{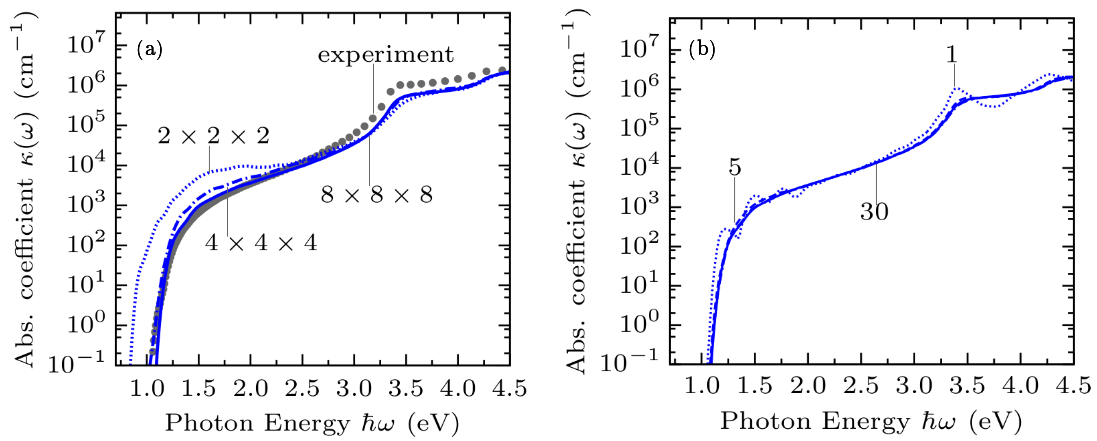}
  \end{figure*}
  \noindent FIG. S6:
  (a) Numerical convergence of the calculated absorption coefficient of silicon with respect 
  to the size of the BVK supercell (blue lines), compared with the experimental data from
  Ref.~[18] (gray filled discs). Dotted, dash-dotted, and solid lines refer
  to 2$\times$2$\times$2, 4$\times$4$\times$4, and 8$\times$8$\times$8 BVK supercells, respectively.
  Larger supercells are expected to further improve the agreement between our calculations and
  experiment at a very fine scale. (b) Numerical convergence of the calculated absorption coefficient 
  of silicon with respect to the sampling of the Brillouin zone. Solid blue lines indicate spectra 
  calculated within a 8$\times$8$\times$8 BVK supercell using 1 (dotted), 5 (dashed), 
  30 (solid) random ${\bf k}$-points.
\newpage

  \begin{figure*}
  \includegraphics[width=0.55\textwidth]{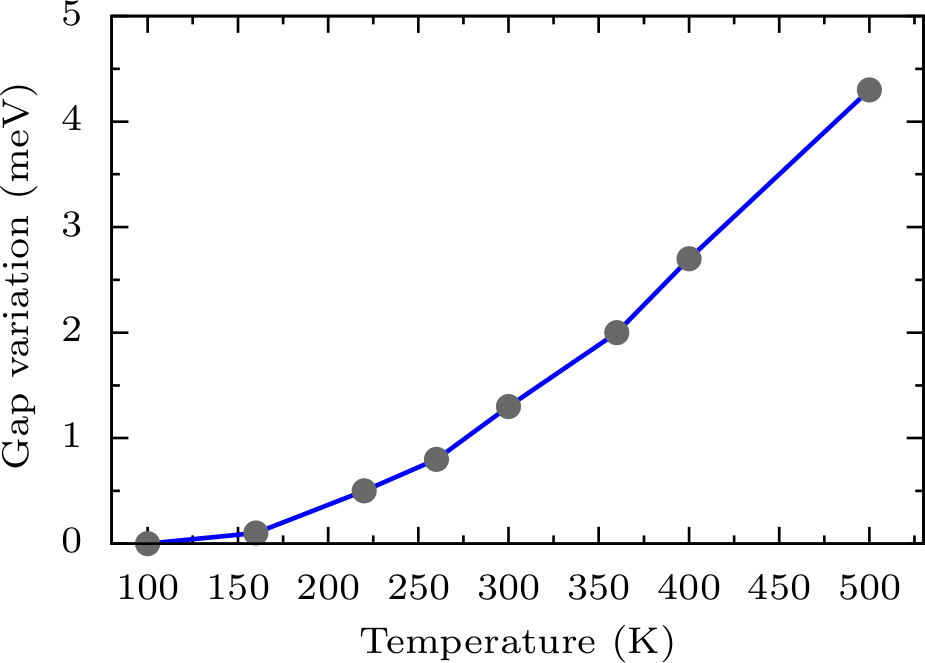}
  \end{figure*}
  \noindent FIG. S7:
Calculated indirect band gap of silicon at clamped nuclei as a function of temperature.
For each temperature we used the experimental lattice parameters from Ref.~[49].
The discs show the change in band gap relative to the value at 100~K, and the lines are
guides to the eye.
As expected the change between 100~K and 500~K is very small ($<5$~meV) as compared to the
variation due to electron-phonon renormalization [$\sim$100~meV, Fig.~2(b)].

\end{document}